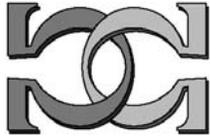
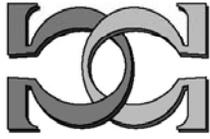

CDMTCS
Research
Report
Series

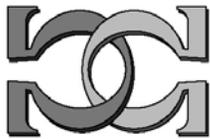
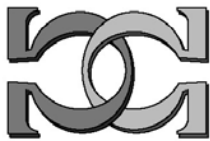

# The Physical World as a Virtual Reality

**Brian Whitworth**
Massey University, Albany, Auckland, New Zealand

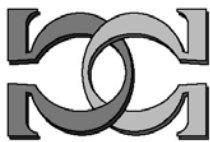

CDMTCS-316
December 2007

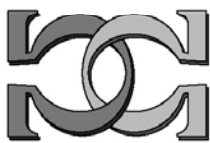
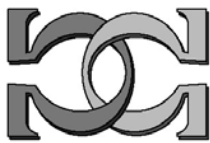

Centre for Discrete Mathematics and
Theoretical Computer Science

# The Physical World as a Virtual Reality

Brian Whitworth
Massey University, Albany, Auckland, New Zealand
E-mail: bwhitworth@acm.org

*Not only is the universe stranger than we imagine, it is stranger than we can imagine*

Sir Arthur Eddington

### Abstract

*This paper explores the idea that the universe is a virtual reality created by information processing, and relates this strange idea to the findings of modern physics about the physical world. The virtual reality concept is familiar to us from online worlds, but our world as a virtual reality is usually a subject for science fiction rather than science. Yet logically the world could be an information simulation running on a multi-dimensional space-time screen. Indeed, if the essence of the universe is information, matter, charge, energy and movement could be aspects of information, and the many conservation laws could be a single law of information conservation. If the universe were a virtual reality, its creation at the big bang would no longer be paradoxical, as every virtual system must be booted up. It is suggested that whether the world is an objective reality or a virtual reality is a matter for science to resolve. Modern information science can suggest how core physical properties like space, time, light, matter and movement could derive from information processing. Such an approach could reconcile relativity and quantum theories, with the former being how information processing creates space-time, and the latter how it creates energy and matter.*

Modern online games show that information processing can create virtual "worlds", with their own time, space, entities and objects, e.g. "The Sims". However that *our* physical world is a virtual reality (VR) is normally considered a topic of science fiction, religion or philosophy, not a theory of physics. Yet the reader is asked to keep an open mind, as one should at least consider a theory before rejecting it. This paper asks if a world that behaves just like the world we live in could arise from a VR simulation. It first defines what VR theory entails, asks if it is logically possible, then considers if it explains known facts better than other theories.

### Strange Physics

While virtual reality theory seems strange, so do other current theories of physics, e.g. the many-worlds view of quantum physics proposes that each quantum choice divides the universe into parallel universes [1], so everything that can happen does in fact happen somewhere, in an inconceivable "multi-verse' of parallel universes. This is a minority view but surprisingly popular. Even relatively main-stream physics theories are quite strange. Guth's inflationary model suggests that our universe is just one of many "bubble universes" produced by the big bang [2]. String theory suggests the physical world could have 9 spatial dimensions, with six of them "curled up" from our perspective. M-theory suggests our universe lies on a three dimensional "brane" that floats in time along a fifth dimension we cannot register [3, p177-180]. The cyclic-ekpyrotic model postulates that we exist in one of two 3D worlds that collide and retreat in an eternal cycle along a hidden extra connecting dimension [4]. Equally strange are the results of modern physics experiments, where time dilates, space curves, entities teleport and objects exist in many places at once, e.g. at the cosmic level:

1. *Gravity slows time*: An atomic clock on a tall building "ticks" faster than one on the ground.
2. *Gravity curves space:* Rays of light traveling around the sun are bent by curved space.
3. *Speed slows time*. An atomic clock on a flying plane goes slower than one on the ground.
4. *Speed increases mass*. As objects move faster, their mass increases.
5. *The speed of light is absolute.* Light shone from a torch on a spaceship moving at 9/10ths of the speed of light leaves the spaceship at the speed of light.

The above statements don't fit our normal reality concepts, yet they have been experimentally verified, e.g. in 1962 one of two synchronized atomic clocks was flown in an airplane for several days while the other stayed stationary on the ground. The result was, as Einstein predicted, less time passed for the clock on the plane. In relativity theory a young astronaut could leave his twin on Earth and return after a year's high speed travel in space to attend his twin brother's 80th birthday. This is not considered a theoretical possibility, but as something that could actually happen. The quantum level of physics introduces even more strangeness:

1. *Teleportation*. Quantum particles can "tunnel", suddenly appearing beyond a barrier they cannot cross, like a coin in a sealed glass bottle suddenly appearing outside it.
2. *Faster than light interaction*. If two quantum particles are "entangled", what happens to one *instantly* affects the other, even if they are light years apart.
3. *Creation from nothing*. Given enough energy, matter can suddenly appear from an "empty" space (where there was no matter before).
4. *Multiple existence*. Light passing through two slits creates a wave interference pattern. The interference continues if photons are shot through the slits *one at a time*, and *regardless of the time delay*. A quantum entity, it seems, can interfere with itself.
5. *Physical effects without causality*. Quantum events like gamma radiation occur randomly, and no physical cause for them has ever been identified.

It is the strange findings of physics experiments that are driving the strange theories of physics.

**Strange theories**

Modern physics began when Maxwell presented his wave equations in 1900 and Einstein suggested special relativity in 1905 and general relativity in 1915. Despite considerable scientific skepticism, these theories met every experimental and logical test their critics could devise. Their predictive success surprised even their advocates, e.g. in 1933 Fermi's formulas pre-discovered the neutrino (a particle with no significant mass or charge) well before nuclear experiments verified it in 1953. Dirac's equations similarly predicted anti-matter before it too was later confirmed. These and other stunning successes have made the theories of quantum mechanics and relativity *the crown jewels of modern physics*. They have quite simply never been shown wrong. Yet, a century later, *they still just don't make sense*. As Kenneth Ford says of quantum theory:

*"Its just that the theory lacks a rationale. "How come the quantum" John Wheeler likes to ask. "If your head doesn't swim when you think about the quantum," Niels Bohr reportedly said, "you haven't understood it." And Richard Feynman … who understood quantum mechanics as deeply as anyone, wrote: "My physics students don't understand it either. That is because I don't understand it.""* [5, p98]

Similar statements could be made of relativity theory's claims that time and space are malleable. For perhaps the first time in the history of any science, the scholars of physics simply don't personally believe what the reigning theories of their discipline are saying. They accept them as mathematical statements that give correct answers, but not as literal world reality descriptions. This is, to say the least, an unusual state of affairs. The problem is not lack of use, as these theories permeate modern physics applications, from micro-computers to space exploration. By some estimates 40% of US productivity derives from technologies based on quantum theory, including cell phones, transistors, lasers, CD players and computers. Yet physicists use quantum theory because it works not because it makes sense:

"*… physicists who work with the theory every day don't really know quite what to make of it. They fill blackboards with quantum calculations and acknowledge that it is probably the most powerful, accurate, and predictive scientific theory ever developed. But … the very suggestion that it may be literally true as a description of nature is still greeted with cynicism, incomprehension, and even anger.*" [6]

The need is not for more proofs or applications but for more understanding. Physicists know the mathematics, but cannot connect it to their practical knowledge of the world, i.e. the theories are useful but not meaningful. Physics has theories that work but which make no sense, e.g. Feynman observed that an electron traveling from A to B acts like it simultaneously traverses all possible intervening paths. His "sum over histories" theory gives the mathematics to do this calculation, and it predicts quantum outcomes well. Yet while most scientific theories increase understanding, this theory seems to take understanding away. How can *one* electron simultaneously travel *all* possible paths between two points? Is the theory just a mathematical device, not a reality description?

It is ironic that relativity theory and quantum theory not only contradict much of what we know (or think we know) of the world, they also contradict each other. Each has its domain - relativity describes macro space-time events, and quantum theory describes micro sub-atomic events. Each theory works perfectly within its own domain, but combining them creates contradictions, e.g. relativity demands that nothing can travel faster than light, but in quantum entangled particles can affect each other instantly from anywhere in the universe, which Einstein called "spooky action at a distance". As Greene notes:

"*The problem … is that when the equations of general relativity commingle with those of quantum mechanics, the result is disastrous.*" [7, p15]

A symptom of the semantic failure of modern theoretical physics is that even after a century of successful use and testing, even simple versions of its main theories are not yet taught in high schools, perhaps as it is difficult to teach what one doesn't believe. Physics has contained the problem by putting a mathematical "fence" around it, perhaps as a sort of quarantine:

"*… we have locked up quantum physics in "black boxes", which we can handle and operate without knowing what is going on inside.* [8] (Preface, p x).

Relativity and quantum theory today have become like magic wands, which physicists manipulate to predict the universe, but why or how the mathematical "spells" work is unknown. Some argue that pragmatically it doesn't matter - if the mathematics works what else is needed? Yet others think that since these formulae describe the essence of physical reality, an explanation is due: "*Many physicists believe that some reason for quantum mechanics awaits discovery.*" [5, p98]

One cannot relegate quantum and relativity effects to the "odd" corner of physics, as in many ways these theories *are* modern physics. Quantum theory rules the atomic world, from which the visible world we see emerges. Special and general relativity rule the cosmic world of vast space, which surrounds and contains our world. Between these two poles, everything we see and know about the physical world is encompassed. It is unacceptable that these theories, however mathematically precise, continue to remain opaque to human understanding. Yet modern physics increasingly describes a physical world in which information is central. Virtual reality theory arises from the Sherlock Holmes dictum: "*…when you have excluded the impossible, whatever remains, however improbable, must be the truth*". Let us now postulate the unthinkable: that the "real" world is a virtual reality.

## The virtual reality axiom

While never commonly held, the idea that the world is a virtual reality has a long pedigree. Over two thousand years ago Pythagoras thought numbers were the non-material essence from which the physical world was created. Buddhism says the world is an illusion, and Hinduism considers it God's "play" or Lila, while Plato's cave analogy suggested the world we see is like shadows on a cave wall, and reflects rather than is reality. Plato also felt that "God geometrizes", and Gauss believed that "God computes" (Svozil, 2005), both arguing that the divine mind appears as nature's mathematical laws. Blake's illustration "The Ancient of Days" shows Urizen wielding a compass upon the world. Zuse first expressed the concept in modern scientific terms, suggesting that space calculates [9], and since then other scientists have also considered the idea [10-16].

A *virtual reality* is here considered to be a reality created by information processing, and so by definition it cannot exist independently in and of itself, as it depends upon processing to exist. If the processing stops then the virtual reality must also cease to exist. In contrast an *objective reality* simply *is*, and does not need anything else to sustain it. This suggests two hypotheses about our reality:

1. **The objective reality (OR) hypothesis:** *That our physical reality is an objective reality that exists in and of itself, and being self-contained needs nothing outside of itself to explain it.*

2. **The virtual reality (VR) hypothesis**: *That our physical reality is a virtual reality that depends upon information processing to exist, which processing must occur outside of itself.*

Whatever one's personal opinion, these views clearly contradict. If the world exists as an objective reality it cannot be virtual, and if it exists as a virtual reality then it cannot be objective. That the world is an objective reality and that it is a virtual one are mutually exclusive. Each hypothesis has implications, e.g. objective reality suggests the universe as a whole is permanent, as it has nowhere to come from or go to. It implies the sort of *physical realism* statements that quantum theory contradicts, for example [17]:

1. *Object locality*: That objects exist in a locality that limits their event interactions.

2. *Object reality*: That objects have inherent properties that their existence carries forward from one moment to the next, and these determine their behavior independent of any measurement.

To illustrate the depth of the contrast, consider the primary axiom of Lee Smolin's recent book:

> "***There is nothing outside the universe***" [18 p17].

The edifice of science itself is often assumed to rest upon this apparently self-evident statement, yet it is precisely this statement that VR theory contradicts. Indeed the prime axiom of virtual reality theory can be obtained by reversing Smolin's axiom, namely:

> ***There is nothing in our universe that exists of or by itself.***

This axiom arises because a VR processor cannot itself logically exist within the virtual reality its processing creates. A processor cannot create itself because the virtual world creation could not start if a processor did not initially exist outside it. Hence any VR world, by definition, *must* have existence dimensions outside itself. Many physics theories, like string theory, already suggest that our world has additional dimensions, yet these are for some reason still assumed to be in the world, but just "curled up" to be invisible to us. In contrast VR theory's additional dimension(s) must be outside the VR world. Yet what is the difference between an unknowable dimension that is "in the world" and one that is "outside the world"? Since both are untestable science favors neither view. To postulate the world is virtual does not contradict science, but rather engages its spirit of questioning. Science is a method of asking questions, not a set of reality assumptions [19]. Scientists are entitled to ask if what could be actually is so. The only constraint is that the question be decided by feedback gathered from the world by an accepted research method. Science does not require an objective world, only information to test theories against, which a VR can easily provide. Not only can science accommodate the virtual world concept, a virtual world could also sustain science.

## Can a virtual reality be real?

Doesn't common sense deny that the world which appears so real to us is a virtual reality? Philosophers like Plato have long recognized that the reality of reality is not provable [20]. Bishop Berkeley's solipsism argued that a tree falling in a wood will make no sound if no-one is there to hear it. Dr Johnson is said to have reacted to that idea the world is created by the mind by stubbing his toe on a stone and saying "I disprove it thus". However VR theory does not claim that the world is unreal to its inhabitants, only that it is not objectively real.

To clarify the difference, suppose information processing in one world creates a second virtual world. To an observer in the first world, events within the virtual world are "unreal", but to an observer within the virtual world, virtual events are as real as it gets. If a virtual gun wounds a virtual man, to that virtual man the pain is "real". That a world is calculated does not mean it has no "reality", merely that its reality is local to itself. Even in a virtual reality, stubbed toes will still hurt and falling trees will still make sounds when no-one is around. Reality is relative to the observer, so by analogy, a table is "solid" because our hands are made of the same atoms as the table. To a neutrino, the table is just a ghostly insubstantiality through which it flies, as is the entire earth. Things constituted the same way are substantial to each other, so likewise what is "real" depends upon the world it is measured from. To say a world is a virtual doesn't imply it is unreal to its inhabitants, only that its reality is "local" to that world, i.e. not an objective reality.

The science-fiction movie The Matrix illustrated how a calculated reality could appear real to its inhabitants (as long as they remained within it). This was possible because people in the matrix only knew their world from the information they received, which is exactly how we know ours. Yet this movie does not illustrate VR theory, as its matrix was created by machines in a physical world, and matrix inhabitants could escape to this "real" world, i.e. the physical world was still presumed to be the "end of the line" for "realness". In contrast VR theory does not assume this. It merely argues that our reality is a local reality, i.e. dependent upon processing outside itself. Yet the Matrix movie did correctly show that a virtual world need not be obviously so:

*"But maybe we are all linked in to a giant computer simulation that sends a signal of pain when we send a motor signal to swing an imaginary foot at an imaginary stone. Maybe we are characters in a computer game played by aliens."* [6, p131]

However Hawking's next sentence was "*Joking apart, ...*" Though logically our world *could* be virtual, for some reason to imagine that it *is* can only be presented as a joke involving aliens.

## Approaching virtual reality

Current physics seems to approach VR theory in three ways:

*The physical world as a virtual reality, Brian Whitworth**The physical world as a virtual reality, Brian Whitworth*

1. **Calculable Universe Hypothesis**: That our physical reality can be simulated by information processing that is calculable (halting).

2. **Calculating Universe Hypothesis**: That our physical reality uses information processing in its operation to some degree.

3. **Calculated Universe Hypothesis:** That our physical reality is created by information processing based outside the physical world we register.

*The calculable universe hypothesis* states that physical reality **can be simulated by information processing** [14]. Calculable here does not mean deterministic, as processing can be probabilistic, nor does it mean mathematically definable, as not all definable mathematics is calculable, e.g. an infinite series. Many scientists accept that the universe is calculable in theory, as the Church-Turing thesis states that for any specifiable output there is a finite program capable of simulating it. If our universe is lawfully specifiable, even probabilistically, then in theory a program could simulate it (though this universal program might be bigger than the universe itself). This hypothesis does not say the universe *is* a computer, but that it could be simulated by one, i.e. it does not contradict objective reality.

*The calculating universe hypothesis* states that the universe **uses information processing** algorithms to create reality, e.g. quantum mechanical formulae. Supporters of this view are a minority, but include mainstream physicists like John Wheeler, whose phrase "*It from Bit*" suggests that objects ("it") somehow derive from information ("bit"). Now information processing does not just *model* the universe, it *explains* it [21]. While a computer simulation *compares* its output to the physical world, in a computer explanation the information processing *creates* reality, i.e. the latter is a theory about how the world actually works. Now the world is not just *like* a computer, it *is* a computer.

*The calculated universe hypothesis* goes a step further, stating that physical reality **is created by** external information processing, which equates to the VR hypothesis presented earlier. Now the physical "real" world is the computer *output* rather the computer *process*. Supporters of this "strong" virtual reality theory are few [10], with none in the physics mainstream.

A common criticism of the calculated hypothesis is that we *"…have no means of understanding the hardware upon which that software is running. So we have no way of understanding the real physics of reality."* [22]. The argument is that virtuality implies an unfalsifiable reality, and so is unscientific and should be dismissed. However this misrepresents VR theory, which postulates no other dimensional "hardware". It is a theory about *this world,* not some other unknowable world, and its hypothetical contrast is that *this world* is an objective physical reality. Unprovable speculations about other virtual universes [23], or that the universe could be "saved" and "restored" [11], or that our virtual reality could be created by another VR [24], fall outside the scope of VR theory as proposed here. Further, the theory that the world is an objective reality is just as unprovable as the theory that it is a virtual reality. It is inconsistent to dismiss a new theory because it is unprovable when the accepted theory is in exactly the same boat.

The above three hypotheses cumulate, as each requires the previous to be true. If the universe is not calculable it cannot use calculating in its operations, and if it cannot operate by calculating it cannot be a calculated reality. Hence VR theory is falsifiable as one could disprove it by showing some incomputable physics. If reality does something that information processing cannot, then the world cannot be virtual, which supports the objective reality hypothesis. Yet while there are many incomputable algorithms in mathematics, all known physics seems to be computable.

The above three hypotheses also constitute a slippery slope, as if one accepts that physical reality is *calculable* then perhaps it is also *calculating*, and if it is calculating, then perhaps it is also

77

*calculated*, i.e. virtual. On the surface the calculating universe hypothesis seems to give the best of both worlds, combining an objective universe and information processing, e.g. Deutsch says:

"*The universe is not a program running somewhere else. It is a universal computer, and there is nothing outside it.*" [25]

Yet if the physical world is a universal computer with nothing outside it, *what is its output*? What is the "output" for example of the solar system? While the brain may input and output information like a computer, most of the world does not [21]. Or if the physical world is the computing output, what is doing the processing? That the universe computes the universe creates a recursive paradox [26]. For physical processing, occurring in the physical world, to create that same physical world is an entity creating itself, which is illogical. A universe can no more output itself than a computer can output itself. The physical universe cannot be both a universal computer and its output. If the physical world is produced by information processing, as the computations of modern physics imply, that processing cannot occur in the physical world, i.e. it must occur elsewhere. Under examination, the calculating universe hypothesis collapses to the calculated universe hypothesis, i.e. to VR theory, giving only two viable theoretical alternatives – objective reality and virtual reality.

**Virtual reality requirements**

If one were to create a VR that behaves like our world, what would be the requirements? To proceed, one must assume ***information processing constancy:*** *that information processing operates the same way in all worlds,* e.g. information processing in our world involves discrete input/output, calculable algorithmic processes, and finite memory and processing, and so it is assumed that virtual reality processing works the same way. Other requirements include:

1. ***Finite processing allocations***. *That the processing that creates a VR that behaves like our world allocates it's processing in finite amounts*. Apart from the fact that we have no concept of what "infinite" processing means, finite processing allocation suggests that every quanta of matter, time, energy and space has a finite information capacity: *"…recent observations favor cosmological models in which there are fundamental upper bounds on both the information content and information processing rate.*" [27 ,p13] While the processing power needed to run a universe is enormous it is not inconceivable, e.g. Bostrom argues that all human history would require less than $10^{36}$ calculations to simulate, and a planet sized computer could provide $10^{42}$ operations per second [24].

2. ***Autonomy***. *Once started, a VR that behaves like our world must run itself without further information input.* Most human computer simulations require regular data input to run. In a virtual world that behaves like ours, such external data input would constitute a "miracle", and in our world miracles are at best rare. This VR simulation must run itself without miracles, i.e. without ongoing data input.

3. ***Consistent self-registration***. *A VR that behaves like our world must "register" itself consistently to internal "observers".* Most human computer simulations output data to an outside viewer, but we see our world from within. We register "reality" when light from the world interacts with our eyes, also in the same world. For a virtual reality to "register itself" as we do, internal interactions must be consistent with respect to each local "observer".

4. ***Calculability***. *A VR that behaves like our world must at all times be calculable*. A finite processing source must ensure that no calculations tend to infinity, e.g. the processing demands of some many body calculations explode to incalculability. Calculability requires a simulation that is guaranteed to avoid such infinities.

These major requirements constrain any VR model of our world. A prima facie case is now presented that such a model could help explain some of the strange results of modern physics.

**A prima facie case that the physical world is a virtual reality**

One of the mysteries of our world is how every photon of light, every electron and quark, and indeed every point of space itself, seems to just "know" what to do at each moment. The mystery is that these tiniest parts of the universe have no mechanisms or structures by which to make such decisions. Yet if the world is a virtual reality, this problem disappears. Other examples of how a VR approach could illuminate current physics issues include:

1. *Virtual reality creation*. A virtual reality usually arises from "nothing", which matches how the big bang theory proposes our universe did arise (see next section).

2. *Maximum processing rate*. The maximum speed a pixel in a virtual reality game can cross a screen is limited by the processing capacity of the computer running it. In general, a virtual world's maximum event rate is fixed by the allocated processing capacity. In our world, the fixed maximum that comes to mind is the speed of light. That there is an absolute maximum speed could reflect a maximum information processing rate (see next section).

3. *Digital processing*. If a world is virtual, everything in it must be digitized, and so discrete at the lowest level. Plank's discovery that light is quantized (as photons) could then generalize not only to charge, spin and matter, but also to space-time. Discrete space-time avoids the mathematical infinities of continuous space-time, as loop quantum gravity theory argues [18].

4. *Non-local effects*. The processing that creates a virtual world is not limited by the space of that world, e.g. a CPU drawing a screen is no "further" from any one part of the screen than any other. All screen points are equidistant with respect to the CPU, so VR processor effects can ignore screen distance, i.e. be non-local. If our universe is a three-dimensional "screen" it's processing is "equidistant" to all points in the universe, so the non-local collapse of the quantum wave function could be such an effect.

5. *Processing load effects*. On a distributed network, nodes with a high local workload will slow down, e.g. if a local server has many demands a video download may play slower than usual. Likewise a high matter concentration may constitute a high processing demand, so a massive body could slow down the information processing of space-time, causing space to "curve" and time to slow. Likewise, if faster movement requires more processing, speeds near light speed could affect space/time, causing time to "dilate" and space to extend. Relativity effects could then arise from local processing overloads.

6. *Information conservation*. If a system inputs no new information after it starts, it must also not lose the information it has or it will "run down". Our universe has not run down after an inconceivable number of microscopic interactions over 14+ billion years, so if it is made of information it must conserve it. If matter, energy, charge, momentum and spin are all information, all the conservation laws could reduce to one. Einstein's transformation of matter into energy ($e=mc^2$) would then be simply information going from one form to another. The only conservation law VR theory requires is that of information conservation.

7. *Algorithmic simplicity*. If the world arises from finite information processing, it is necessary to keep frequent calculations simple. Indeed the core mathematical laws that describe our world are surprisingly simple: "*The enormous usefulness of mathematics in the natural sciences is something bordering on the mysterious and there is no rational explanation for it*." [28] In VR theory physical laws are simple because they must actually be calculated.

8. *Choice creation*. Information arises from a choice between options [29]. A mechanical or predictable choice is not really a choice in this sense. Einstein never accepted that quantum events were truly random, i.e. no prior world events could predict them. That a radioactive atom decays by pure chance, whenever "it decides" was to him unacceptable, as it was a physical event not predicted by another physical event. He argued that one day quantum random effects would be predicted by as yet unknown "hidden properties". Yet if the source of quantum randomness is the VR processor, which is outside the physical world, this predicts that no hidden variables will ever be found.

9. *Complementary uncertainty*. In Newtonian mechanics one can know both the position and momentum of objects, but for quantum objects Heisenberg's uncertainty principle means one cannot know both at once. Knowing one property with 100% certainty makes the other entirely uncertain. This is not measurement "noise", but a property of reality, e.g. measuring particle position displaces its momentum information, and vice-versa. In a similar way virtual reality "screens" are typically only calculated when they are viewed, i.e. when an interaction occurs [12]. If complementary object properties use the same memory location, the object can *appear* as having either position *or* momentum, but not both at once.

10. *Digital equivalence*. Every digital symbol calculated by the same program is identical to every other, e.g. every "a" on this page identical to every other one because all arise from the same computer code. In computing terms, objects can be "instances" of a general class. Likewise every photon in the universe is exactly identical to every other photon, as is every electron, quark, etc. While the objects we see have individual properties, quantum objects like photons seem all pressed from identical moulds. VR theory suggests that this is so because each is created by the same digital calculation.

11. *Digital transitions*. When one views a digital animation it looks continuous, but in fact it is a series of state transitions, e.g. a movie is a series of still frames run together fast enough to look like a continuous event. Yet if the projector is slowed down, one sees a series of still pictures. Quantum mechanics describes quantum interactions in similar terms, as state transitions. These transitions could explain quantum tunneling, where an electron at A suddenly appears at C without moving through the intervening area B which is impenetrable to it. While this is strange for an objective reality, in VR theory all object movement would be expected to be by state transitions.

Individually none of the above short points is convincing, but taken together they constitute what a court might call circumstantial evidence, favoring virtual reality against objective reality. When coincidences mount up, they present a plausibility argument if not a proof. More powerful evidence is provided by cases which a VR theory explains easily but which OR approaches have great difficulty with. Two such cases are now given in more detail.

### Where did the universe come from?

The traditional view of our universe was that as an objective reality it "just is", and so has always existed. While its parts may transform, its total is in a "steady state" that always was and always will be. The alternative view is that the universe did not always exist, but arose at some specific point, which also created space and time. During the last century these two theories have battled it out for supremacy on the stage of science. Steady-state theory proponents included respected physicists, who thought that the idea that the entire universe expanded from a single point was highly unlikely to be true. However Hubble's finding that all the stars around us are red-shifted suggested that the entire universe is indeed expanding at the speed of light. Now an expanding universe has to expand from somewhere, so scientists could run the expansion backwards to a source, a "big bang" that began our universe about 15 billion years ago. The discovery of cosmic background radiation, left over from the big bang, has largely confirmed the theory today in the minds of most physicists.

Big bang theory sidesteps questions like: "What existed before the big bang?" by answering: "There was no time or space before the big bang", but if time and space suddenly "appeared" for no apparent reason at the big bang, could they not equally suddenly disappear tomorrow? Big bang theory implies a dependent universe, so what is it dependent upon is a valid question even without time and space. If nothing in our universe is created from nothing, how can an entire universe come from nothing? That our universe arose from nothing is not just incredible, it is inconceivable. One can state the problems simply:

1. What caused the big bang?
2. What caused space to start?
3. What caused time to start?
4. How can a big bang arise when there is no time or space?
5. How can space be caused if there is no "there" for a cause to exist within?
6. How can time be started if there is no time flow for the starting to occur within?

The big bang contradicts any theory that assumes the universe is objectively real and complete in itself. How can an objective reality, existing in and of itself, be created out of nothing? The failure of the steady state theory of the universe removes a cornerstone of support for the objective reality hypothesis. In contrast virtual reality theory fits well with a big bang. No virtual reality can have existed forever, since it needs a processor to start it up. All virtual realities "start up" at a specific moment of time, typically with a sudden influx of information. Every time one starts a computer game or boots up a computer, such a "big bang" occurs. From the perspective of the virtual world itself, its creation is always from "nothing", as before the virtual world startup there was indeed no time or space *as defined by that world*. There was nothing relative to that world because the world itself did not exist. It is a hallmark of virtual realities that they must come into existence at a specific event in their space and time, which also initiates their space-time fabric. Note that in a virtual world there is no logical reason why all initiating information cannot initially "point" to a single arbitrary location, i.e. no reason why an entire universe cannot exist at a single point. In VR theory the big bang was simply when our universe was "booted up".

The big bang is an accepted aspect of modern physics that VR theory accommodates but OR theory does not. It illustrates that VR/OR arguments can be resolved by appeal to experimental data *from this world*. Just as the steady state versus big bang theories were resolved by research, so can the more general virtual vs. objective theoretical contrast be resolved. To decide if the world is objective or virtual we simply need to consider what data from the world is telling us.

## Why does our universe have a maximum speed?

The author's interest in this subject began with a simple question: *"Why does our universe have a maximum speed?"* Einstein deduced that nothing travels faster than light from the way the world works, but this did not explain why the world had to be that way. Why cannot an object's speed simply keep increasing? Why must there be a maximum speed at all? If light is like a classical wave, its speed should depend upon the elasticity and inertia of the medium it travels through. If light travels through the medium of empty space, its speed should depend upon the elasticity and inertia of space. However how can empty space have properties? Once space was considered a luminiferous "ether", through which objects move as a fish swims through water. However such a space would give a fixed frame of reference to movement, and in 1887 Michelson and Morley showed that space didn't work that way. When Einstein deduced that the speed of light was the real absolute, this discredited the spatial "ether" idea.

However this left a problem, namely that empty space (the medium that transmits light) was "nothing". Mathematical properties of empty space, like length, breadth and depth, give no basis for elasticity or inertia. How can the properties of a "nothing" vacuum imply a maximum speed? To say the speed of light defines the elasticity of space argues backwards, that an outcome determines a cause. The nature of space should define the rate of transmission through it. The speed of light should conclude the argument, not begin it. Yet if "empty space" is devoid of object properties, how can it be a "medium" that not only transmits light but also limits its speed?

This paradox, like many others, arises from assuming that there is an objective reality. If one assumes objects exist in and of themselves one must also assume a context for them to exist within. The ether's proponents assumed space (the context) was an "object" like the objects it contained, as both fish and water are physical objects. Einstein showed that space, which contains objects, cannot also itself be an object, else it would exist in itself, which is impossible. Yet Einstein replaced space and time by an equally absolute space-time concept:

*"…absolute space-time is as absolute for special relativity as absolute space and absolute time were for Newton …"* [7, p51]

Einstein replaced the old object context (space/time) with a new context (space-time), but it was still a context. Like Newton, he believed that *objects exist of themselves,* which is what put him at odds with quantum theory's non-local equations. Any theory that assumes objects exist independently must also *assume* a reality context for them to exist within. Such an assumed context, whether space or space-time, cannot have properties like the objects it contains. Yet the speed of light limit suggests that space as a medium of transmission does have properties. String theory has the same problem, as strings are assumed to exist in a space-time context. In contrast virtual reality theory assumes nothing except that everything is information. While objective reality must assume space, time, or both, virtual reality theory does not.

Information, as a universal constituent, avoids the problem that a substance cannot exist within itself because information processing can "stack", i.e. processing can create processing. That VR objects arise from information processing does not conflict with space itself arising the same way. That a virtual space is empty of "objects" then need not make it empty of structure, just as an idle computer network still has protocols and connections to maintain. Space as a virtual processing network supports the modern view that empty space is not "empty". It also allows a maximum network processing rate property. The Lorentz transformations suggest the maximum rate objects can move through space-time is a trade off between space and time, so for a photon moving at the maximum speed of light the rate of change of time is zero, i.e. time stands still. If both space and time arise from a fixed information processing allocation, that the sum total of space and time processing adds up to the local processing available is reasonable. That our universe has maximum change rates is a fact of physics VR theory explains well but objective reality cannot.

## Evaluating virtual reality theory

Possible responses to this prima facie case for the world as a virtual reality include:

1. *Spurious*. One can satisfy the requirements of any world by appropriate assumptions, so a VR model can always be found to match our world. This response is less likely if the model's assumptions are few and reasonable.

2. *Coincidence*. The matches between VR theory and modern physics are fortunate coincidences. This response is less likely if the matches found are many and detailed.

3. *Useful*. Seeing the world in information processing terms may open up new perspectives in physics. This response is more likely if VR theory explains many things.

4. *Veridical*. Our world is in all likelihood a virtual reality. This option is more likely if VR theory explains what other theories cannot.

While it the reader can decide their own response, it is suggested that virtual reality theory is a logical option that deserves consideration alongside physic's other strange theories. That the essence of the universe is information may not be correct, but it is a useful approach to some of the perennial issues of physics. Can science evaluate if a world is a virtual reality from within it? Suppose one day that the computer code that creates "The Sims", a virtual online world, became so complex that some Sims within the simulation began to "think". Could they deduce that their world was a virtual world, or at least that it was likely to be so? If simulated beings in a simulated world acquired thought, like us, would they see their world as we see ours now? A virtual entity could not *perceive* the processing that creates its world, but it could *conceive* it, as we do now. They could compare how a virtual reality would behave with how their world actually behaved. They could not "know", but they could deduce a likelihood, which is all our science does anyway.

Science warns against selecting data to support a theory. It requires unbiased data, not data selected by the researcher (to fit their case). It is not enough to find that *selected* computer programs, like cellular automata, mimic *selected* world properties [13], as the researcher can then choose what is to be explained. There is no need for "a new kind of science" if the old kind still works, i.e. one must not select the parts of reality one's VR theory explains. One way to avoid this trap is to *derive the core of physics from first principles*, i.e. begin with the properties of computing and derive properties like space, time, light, energy, electrons, quarks and movement. This would explain not just selected world events but its operational core. This approach, to assume VR theory is true then "follow the logic" until it fails, has so far been surprisingly successful, as a following paper will show. If the world is not a virtual reality, assuming it is so should soon generate outcomes inconsistent with observations, but if the world is indeed a virtual reality, it should consistently explain facts that objective reality theories cannot. Ultimately, the success or failure of the VR model depends upon how well it explains our world.

## Discussion

Almost a century ago Bertrand Russell dismissed the idea that life is a dream using Occam's razor (that a simpler theory is always preferred):

"*There is no logical impossibility in the supposition that the whole of life is a dream, in which we ourselves create all the objects that come before us. But although this is not logically impossible, there is no reason whatever to suppose that it is true; and it is, in fact, a less simple hypothesis, viewed as a means of accounting for the facts of our own life, than the common-sense hypothesis that there really are objects independent of us, whose action on us causes our sensations*." [30]

However in VR theory objects could be independent of us but still not objectively real. It suggests that all physical entities, all events acting upon them, and the context of space-time itself, arise from information processing. That information is the basic underlying "stuff" of the universe is today not so easily dismissed. Given the big bang, what is simpler, that an objective universe was created out of nothing, or that a virtual reality was booted up? Given the speed of light is a universal maximum, what is simpler, that it depends on the properties of featureless space, or that represents a maximum network processing rate? Similar questions can be asked for each of the points summarized in Table 1. Modern physics increasingly suggests that virtual reality is a simpler theory, i.e. that Occam's razor now favors virtual reality over objective reality.

VR theory does not change the mathematics of physics, but it drastically changes its meaning, as *if the universe is virtual then so are we.* This reduces us to pixilated avatars in a digital world, which hardly flatters the human ego, but then again, science has done this before:

"*Since our earliest ancestors admired the stars, our human egos have suffered a series of blows.*" [14]

Copernicus first discovered that the Earth is not the center of the universe, and we now know that our tiny planet circles a mediocre star two-thirds of the way out of a million, million star galaxy, itself within a million, million galaxy universe. Darwin discovered that we are not the center of things biologically either, since over 99.9% of every species that ever lived are now extinct. Even the matter we are made of is only about 4% of the universe, with the rest being dark matter (23%) and dark energy (73%) [5, p246]. Freud found that the sub-conscious has more impact than the conscious, and neuroscientists find the brain "split" at the highest (cortical) level [31], suggesting our unitary "self" is also an illusion [32]. Science may be preparing further disillusionments in areas like dreams, genetics and consciousness. The trend is clear: science finds us to actually be less than we imagine, and we imagine ourselves to actually be more than science finds we are. Would one more ego blow, say that our reality didn't exist objectively at all, be a surprise?

For a century physicists have tried unsuccessfully to interpret quantum and relativity theories with traditional objective reality concepts. Quantum experiments on Bell's theorem flatly contradict both the locality and reality assumptions of physical realism [17]. It is time to try something new. Yet even physicists who call for radical new views of reality balk at the idea of virtual reality. Modern physics implies a calculated world, but that such a world is the case seems to offend. Yet that we cannot imagine something is so, or that we would wish it were not so, are not reasons for it to actually be not so. Ultimately, whether our world is virtual or real is not our choice, as we must accept our reality whatever form it takes.

Theoretical physics is currently in a conundrum. On the one hand, mathematical speculations about unknowable dimensions, branes and strings seem increasingly pointless and untestable [33]. On the other hand, objective realism seems to face paradoxes it can never, ever, solve. This paper applies computer knowledge to physics, and proposes virtual reality theory as *a real hypothesis about the knowable world*. This approach could open up new ideas, as virtual objects need no inherent properties or locations beyond those embodied in the calculations that create them. A virtual reality theory could reconcile the contradiction between relativity and quantum theory, as the former could be how information processing creates space-time, and the latter how it creates energy, matter and charge. It could also solve the quantum measurement problem, as if our reality is in effect a processing interface, an observer viewing an object could indeed create it. Similarly in an online virtual world the entire world is not calculated onscreen at once. The computer, for practical reasons, only calculates what the viewer chooses to view after they choose to view it, i.e. screen calculations are as required. If what we call reality is a multi-dimensional space-time interface, it would likewise be expected to be calculated only on demand. The virtual reality viewer would then be no more aware of this than a virtual game player is, as everywhere they looked the world would "exist". Our reality could indeed be only calculated when we "measure" it. However there is a twist, as if our world is a virtual reality, we are viewing it from within not without. In a computer game, the player exists outside the screen interface. However in the case of our world, we are viewing it from within. This makes this world a recursive interface, that both sends to and receives from itself. If so, it is like no other information interface that we know.

**Table 1. Virtual properties and physical outcomes**

| Virtual Property | Physical Outcome |
|---|---|
| *Virtual reality creation.* Virtual worlds must begin with an information influx from "nothing", that also begins VR time/ space. | *The big bang.* The universe was created out of nothing by a "big bang" in a single event that also created time and space. |
| *Digital processing.* All events/objects that arise from digital processing must have a minimum quantity or quanta. | *Quantum minima.* Light is quantized as photons. Matter, energy, time, and space may be the same, i.e. have a minimum amount. |
| *Maximum processing rate.* Events in a VR world must have a maximum rate, limited by a finite processor. | *Light speed.* The speed of light is a fixed maximum for our universe, and nothing in our space-time can move faster. |
| *Non-local effects.* A computer processor is equidistance to all screen "pixels", so its effects can be "non-local" with respect to its screen. | *Wave function collapse.* The quantum wave function collapse is non-local - entangled photons on opposite sides of the universe may instantly conform to its requirements. |
| *Processing load effects.* If a virtual processing network is overloaded, its processing outputs must be reduced. | *Matter and speed effects.* Space curves near a massive body and time dilates at high speeds. |
| *Information conservation.* If a stable VR is not to gain or lose information it must conserve it. | *Physical conservation.* Physical existence properties like matter, energy, charge, spin etc are either conserved or equivalently transform. |
| *Algorithmic simplicity.* Calculations repeated at every point of a huge VR universe must be simple and easily calculated. | *Physical law simplicity.* Core physical processes are describable by relatively simple mathematical formulae, e.g. gravity. |
| *Choice creation.* A random number function in the VR processor could provide the choices needed to create information. | *Quantum randomness.* The quantum "dice throw" is to the best of our knowledge truly random, and unpredictable by any world event. |
| *Complementary uncertainty.* Calculating one property of a self-registering interface may displace complementary data. | *Heisenberg's uncertainty principle.* One cannot know both a quantum object's position and momentum, as knowing either makes the other unknown. |
| *Digital equivalence.* Every digital object created by the same code is identical. | *Quantum equivalence.* All quantum objects, like photons or electrons, are identical to each other. |
| *Digital transitions.* Digital processes simulate event continuity as a series of state transitions, like the frames of a film. | *Quantum transitions.* Quantum mechanics suggests that reality is a series of state transitions at the quantum level. |

## Acknowledgements

Thanks to Professor Onofrio L. Russo, NJIT, for first arousing my interest in this subject, and to Professor Ken Hawick, Massey University, for listening to my ramblings.